\documentclass[preprint,floatfix,superscriptaddress,amsmath,amssymb,prl]{revtex4}
\usepackage{mathrsfs}
\usepackage{amssymb}
\usepackage{graphicx}
\usepackage{hyperref}
\usepackage{ulem}
\usepackage{tabularx}
\usepackage{array}
\usepackage{color}
\begin{document}


\title{   Spin Operators for Massive Particles}

\newcommand{\ket}[1]{\vert#1\rangle}
\newcommand{\bra}[1]{\langle#1\vert}
\newcommand{\e}[0]{\mathrm{e}}
\newcommand{\projector}[2]{\vert #1 \rangle\hspace{-0.3ex} \langle #2 \vert}

\author{Taeseung Choi}
 \email{tschoi@swu.ac.kr}
\affiliation{Division of General Education, Seoul Women's University, Seoul 139-774, Korea}
\affiliation{School of Computational Sciences, Korea Institute for Advanced Study, Seoul 130-012, Korea}
 \author{Sam Young Cho}
 \email{sycho@cqu.edu.cn}
\affiliation{Centre for Modern Physics and Department of Physics,
 Chongqing University, Chongqing 400044, China}

\begin{abstract}
Since the discovery a century ago, spin describing the intrinsic
angular momentum of massive elementary particles has exposed
its nature and significant roles in wide ranges of
(relativistic) quantum phenomena and practical applications for
future quantum technology. Emerging
inconsistencies
have also disclosed its telltale incomplete description.
Finding relativistic spins (operators) of massive particles
is a long-standing fundamental problem from the beginning of relativistic quantum mechanics.
Here we present the rigorous derivation and the representation of spin operators
from the spacetime symmetry.
The covariant parity operation, defined by the spin operators, naturally leads to
a fundamental equation equivalent to the covariant Dirac equation,
which manifests existent relativistic spins.
Proper understanding position operator in the Dirac theory on
account of the spin operator through total angular momentum predicts
no Zitterbewegung as well as conserving orbital and spin currents.
The spin operators can be applicable for unraveling the
inconsistencies and for exploring unveiled physics of massive
particles.

\end{abstract}
\maketitle

\section{ Introduction.}
\noindent
 Spin of a massive particle (e.g., electron) has become a very familiar
 and indispensable physical quantity
 in fundamental physics and applied sciences as well as quantum technologies
 \cite{Milestone}
 since it was introduced to explain the broadening of the sodium D-lines
 observed by Zeeman in 1897 \cite{Zeeman}
 and
 the splitting of the silver beam observed by Stern and Gerlach in 1922
 \cite{SGexperiment}.
 Still its crucial roles have been revealed in various quantum phenomena
 such as Kondo effects  \cite{Gordon},
 spin Hall effects \cite{Kato}, quantum spin fluid \cite{Xu},
 spin Hall insulator \cite{Konig},
 quantum entanglements \cite{Hasegawa} and so on.
 Quantum properties of spin have been widely used for realizations of
 spin-based quantum computing \cite{Kane}, qubits \cite{Trauzettel},
 gating and logic operations \cite{Khajetoorians,Zu},
 data storage \cite{Chappert},
 and electronics (spintronics) \cite{Awschalom}
 including spin pump \cite{Cho,Gerardot},
 spin filter \cite{Folk}, and
 fluid spintronics \cite{Takahashi}.
 In spite of such remarkable progresses,
 counterintuitive incongruities have emerged,
 for instance,
 in defining reduced spin state (spin entropy)
 \cite{Peres,Czachor,Saldanha1,Taillebois}
 and
 spin current \cite{Rashba,Sun1,Shi,An},
 and in dealing with spin-dependent forces
 \cite{Zawadzki}
 when a relativistic situation or effect (e.g., spin-orbit coupling) is considered.
 In relativistic quantum information and communication, that is, the severe controversy has occurred whether
 the spin entropy determined by the reduced density matrix
 for the spin is valid
 \cite{Peres,Czachor,Saldanha1,Taillebois}.
 In spintronics,
 it seems that the spin current is not conserved in the presence of spin-orbit coupling
 \cite{Rashba,Sun1,Shi,An} in both relativistic and nonrelativistic situations,
 and that the spin transverse force for a relativistic electron does not exist \cite{Zawadzki}.
 In addition, in particle physics,
 it seems also that the nucleon spin is not simply made up by the quark spins,
 contrary to our desirous belief,
 because experimental observations of the proton's spin reveal an incredibly little contribution
 of quarks on it,
 which is called the proton spin crisis \cite{proton1,Jaffe,Aidala}.
 Such puzzling inconsistencies meet at an underlying common ground, i.e.,
 the origin of (relativistic) spin.
 Even the Zeeman interaction between the nonrelativistic spin
 (the Pauli spin operator) and an external magnetic field,
 being widely used in nonrelativistic quantum phenomena,
 to our best knowledge,
 still does not have a concrete theoretical verification for its origin
 because, although the Dirac equation predicts the correct electron magnetic moment
 \cite{Schwartz},
 the spin itself has not been identified explicitly from the first principles deriving the Dirac equation.
 Indeed, since the birth of Dirac theory in 1928 \cite{Dirac},
 Schr\"odinger \cite{Schrodinger} in 1930 suggested
 the odd oscillating (quivering) motion
 of a free spin-$1/2$ massive (Dirac) particle, i.e., the so-called
 \textit{Zitterbewegung} for an explanation of spin.
 On the other hand,
 many different (relativistic) spin operators have been defined for
 a complete description of spin
 \cite{Pryce,NW,FW,Frenkel,Chakrabarti,Gursey,Bogolubov,Choi13}.
 However, the proposed spins do not provide clear answers on the most fundamental
 questions,
 for examples, how the Dirac equation, as the most successful
 description of massive spin-$1/2$ elementary particles,
 can predict the correct electron magnetic moment,
 where spin comes from for massive particles,
 and how and why handedness of massive particles exists to connect to spin.
 Even leaving the emerging inconsistencies aside, such undoubtedly challenging fundamental problems
 have by now remained as an inherent obstacle as ever
 \textit{ab initio} from the early days of spin.
 The matter of fact is that
 (relativistic) spin operators for massive elementary particles have been
 undiscovered \cite{Bauke}.
 In this article,
 we derive and obtain the two spin operators, whose squares are the second Casimir invariant of the Poincar\'e group,
 for massive elementary particles with any integer or half-integer spin
 from the spacetime symmetry by using the minimal physical requirements.
 We find that each of the two spin operators is responsible for each handedness of massive elementary particles and
 then the corresponding natural representations of the Poincar\'e group are given by either right-handed or left-handed
 representations.
 Both the two spin operators are shown to be the generators of little groups of the Poincar\'e group.
 As a natural consequence of parity operation on
 the direct sum $(1/2,0)\oplus(0,1/2)$ representation,
 a fundamental equation in terms of the spin operator
 is obtained and found to be equivalent to the covariant Dirac equation,
 which shows manifestly the existence of relativistic spin operators
 and why the Dirac equation successfully describes spin-$1/2$ elementary particles
 and their spin magnetic moments.
 We show that the Schr\"odinger's Zitterbewegung is not a physical motion of free Dirac particles
 by understanding proper position operator in the Dirac theory on account of the spin operators.
 Finally, from Noether theorem, we show that the spin is a fundamental conserved
 quantity and obtain the expression of the conserved spin current.

\section{Results}

\noindent\textbf{Derivation of spin operator from the spacetime
symmetry.}
  Our four-dimensional free spacetime is believed to have an
  apparent symmetry that includes translation invariance and Lorentz
  invariance.
  The group of translations and Lorentz transformations is called
  Poincar\'e group.
  In 1939, Wigner classified elementary particles
  by an irreducible unitary representation
  of the Poincar\'e group \cite{Wigner}.
 Massive particles with arbitrary spin are then considered
 as unitary irreducible representations of the Poincar\'e group.
  Yet, in the modern paradigm of elementary particles, which is quantum
  field theory,
  fields are in general non-unitary.
  With the unknown origin of spin,
  such a discrepancy might be also responsible for the inconsistencies.
  It has then been required deeper understanding and reinvestigating
  the irreducible representations of Poincar\'e group
  to embed especially elementary particles into fields.
 To find massive particles' spins as the fundamental quantity,
 we start with the Poincar\'e group.

 The most rigorous way to represent a group is to use Casimir operators that
 commute with all generators of the group.
 Essentially,
 the two invariant Casimir operators of the Poincar\'e group,
 i.e.,
 $P^\mu P_\mu$ with the eigenvalue $p^\mu p_\mu=m^2$ and $W^\mu W_\mu$ with
 the eigenvalue $w^\mu w_\mu=-m^2 s(s+1)$, are known to give
 the mass $m$ and the spin $s$ of the particle,
 respectively, where  the Pauli-Lubanski (PL) vector is defined as
 $W^\mu = \frac{1}{2} \epsilon^{\,\mu\nu\rho\sigma} J_{\nu\rho} P_\sigma$
 with a four-dimensional Levi-Civita $\epsilon_{\,\mu\nu\rho\sigma}$
 (we set $\epsilon_{0123}=\epsilon^{1230}=1$),
 the generators of the (homogeneous) Lorentz group $J^{\mu\nu}$,
 and the generators of translations $P^\mu$.
 Here, Einstein summation convention is used for the Greek indexes $\mu \in \{0,1,2,3\}$
 and will be also used for Latin indexes $k\in \{1,2,3\}$,
 unless otherwise specifically stated.
 We will omit the word `operator' freely, e.g.,
 PL vector instead of PL vector operator, because
 the context will clarify the usage.
 The metric tensor $g_{\mu\nu}= \mathrm{diag} (+,-,-,-)$ will be used.
 However, as is known, the spatial components of the PL vector $W^k$
 cannot be a spin three-vector because they do not satisfy
 even the basic requirement of a spin operator,
 i.e., the $\mathfrak{su}(2)$ algebra.
 Then spin operators were not identified explicitly although the PL vector
 reveals spin quantum numbers for massive particles.

 However, the Lorentz-invariant square of the PL vector
 offers a way to reach the proper spin operators satisfying the $\mathfrak{su}(2)$ algebra,
 as the generators of an $\mathrm{SU}(2)$ subgroup
 of the Poincar\'e group, for massive elementary particles with spin $s$.
 The second Casimir invariant actually implies that
 the square of (relativistic)
 spin three-vector ${\bf S}$ is well-defined
 in the Poincar\'e group if it satisfies
 \begin{equation}
  {\bf S}\cdot {\bf S} = -\frac{1} {m^2} W^\mu W_\mu ,
  \label{C2}
 \end{equation}
 that is, the Casimir operator is ${\bf S}^2$
 and on an irreducible representation, ${\bf S}^2$ is equal to $s(s + 1)$ times the
 identity matrix, with $s = 0, \frac{1}{2}, 1, \cdots$.
 Practically, this fact allows us to consider a spin three-vector ${\bf S}$
 as a linear combination of PL vectors
 and to find the explicit expression
 of spin three-vector from physical requirements.
  Thus a general form of spin three vector
  (the $k$-component of ${\mathbf S}$) can be written down as
 \begin{equation}
  S^k = a_{k,0} W^0 + a_{k,k} W^k + a_{k,m\neq k} W^m
     \equiv a_{k,\mu}\ W^\mu,
  \label{eq:linear}
 \end{equation}
 where the index $k$ in $a_{k,k}$ is not considered as repeated.
 The coefficients $a_{k,\mu}$
 will be determined by physical requirements.
 Since the momentum and spin operators
 construct the two independent Casimir operators,
 the momentum and the spin
 (an eigenvalue of the $S^k$) are expected to label
 the representation of the Poincar\'e group.
 This requires that the $S^k$ should commute with the momentum operator $P^\mu$.
 Then the coefficients $a_{k,\mu}$ are functions of complex numbers
 and the momentum operators $P^\mu$
 but they are not functions of the Lorentz generators $J_{\mu\nu}$.
 Since a spin three vector is given
  from the components of the dual spin tensor $^*S^{\mu\nu} = \frac{1}{2}\epsilon^{\mu\nu\rho\sigma}S_{\rho\sigma}$, i.e.,
  $^*S^{k0} = S^k =\frac{1}{2} \epsilon_{kij}S_{ij}$,
  equation (\ref{eq:linear}) should satisfy tensorial properties in spacetime,
  where $\epsilon^{k0ij}$ becomes the three-dimensional Levi-Civita $\epsilon_{kij}$ with $\epsilon_{123}=1$.
 Actually, we find the minimal requirements
 determining the coefficients $a_{k,\mu}$, that is,
 a spin operator should satisfy (i) the $\mathfrak{su}(2)$ algebra
 and (ii) the Lorentz-transformation properties as a second-rank spin tensor.
 The two spin operators based on these physical conditions are obtained (Methods) as
 \begin{equation}
 S^k_{\pm} = \frac{1}{m^2} \left( P^0 W^k - P^k W^0\right)
 \pm \frac{i}{m^2} \epsilon_{0kml}  P^l W^m .
 \label{eq:pm}
 \end{equation}
 The $S^k_\pm$ in equation (\ref{eq:pm}) are valid for all reference
 frames because the $S^k_\pm$ are functions of frame-independent
 operator.
 Note that there exist the two spin
 operators, which will give clear answer on the fundamental questions
 and provide more profound understanding on elementary particles with spin.

%

 To be a proper spin operator in the Poincar\'e group,
 the derived spin operators $S^k_\pm$ in equation (\ref{eq:pm})
  must give the second Casimir invariant $W^\mu W_\mu$.
 Straightforwardly,
 one can show $S^k_\pm S^k_\pm = - W^\mu W_\mu/m^2 $.
 In fact, the two spin operators offer the same Casimir operator
 of the Poincar\'e algebra, i.e., $S^k_+ S^k_+ = S^k_- S^k_-$.
 The $S^k_+ S^k_+$ and $S^k_- S^k_-$
 have the eigenvalues $s_+(s_+ + 1)$ and $s_-(s_-+1)$, respectively,
 because the $S^k_\pm$ are the generators of $\mathrm{SU}(2)$ groups.
 The two spin operators do not commute each other, i.e., $[S^i_+, S^j_-]\neq 0$
 and cannot be mapped to each other by a similarity transformation.
 Consequently,
 there are two inequivalent representations
 for a massive particle with mass $m$ and spin $s_\pm$.
 Actually,
 the two representations are associated with the transformation properties
 of particle's states
 under the Lorentz boost transformations, i.e., particle's handedness.
 It will become clear in detailed discussions of the following discussions.

\noindent\textbf{Representations of Poincar\'e group for the two spin operators
 and particle's handedness.}
 All representations of the Poincar\'e group are classified by the eigenvalues
 of two Casimir invariants,
 $P^\mu P_\mu = m^2$ and $S^k_\pm S^k_\pm = s_\pm (s_\pm +1)$.
 The base states $\Psi_{\pm}(p^\mu,\lambda_\pm)$
 of a representation space $(m,s_\pm)$, on which the representation of the Poincar\'e group acts, are obtained by
 the following eigenvalue equations:
\begin{subequations}
\begin{eqnarray}
 P^\mu\ \Psi_{\pm}(p^\mu,\lambda_\pm) &=& p^\mu\, \Psi_{\pm}(p^\mu,\lambda_\pm),
 \label{eq:psk}
 \\
 S^k_\pm\  \Psi_{\pm}(p^\mu,\lambda_\pm) &=& \lambda_\pm \Psi_{\pm}(p^\mu,\lambda_\pm),
\label{eq:sk}
\end{eqnarray}
\end{subequations}
 where
 $p^\mu=(p^0,\mathbf{p})$, i.e., $\mathbf{p}$ is the spatial momentum
 of the base states
 in a specific frame moving
 with velocity $-\boldsymbol{\beta}= -{\bf p}/ m\gamma $ with respect to
 $k^\mu = (m, \mathbf{0})$,
 We will call $k^\mu = (m, \mathbf{0})$ the particle rest frame (PRF)
 for simplicity and $p^\mu = (p^0, \mathbf{p})$ a moving frame with momentum $\mathbf{p}$.
 Here, the Lorentz factor is $\gamma=1/\sqrt{1- {\boldsymbol{ \beta}^2}}$
 and $\lambda_\pm \in \{-s_\pm, -s_\pm+1, \cdots, s_\pm\}$
 are the spin eigenvalues of the $k$-component of
 the spin operators $\mathbf{S}_\pm$. We use the natural unit $c=\hbar=1$.
 Note that the upper case and the lower case letters $P^\mu$ and $p^\mu$ are used
 for the momentum operator and the momentum eigenvalue, respectively.
 After the momentum operators in the spin operators $S^k_\pm$ act on
 the momentum eigenstate,
 the spin operators in the specific frame moving with
 the momentum $p^\mu$
 become the $S^k_\pm (\mathbf{p})$ satisfying
\begin{eqnarray}
\label{eq:sk}
 S^k_\pm (\mathbf{p})  \Psi_\pm(p^\mu,\lambda_\pm)
 =\lambda_\pm  \Psi_\pm(p^\mu,\lambda_\pm).
\end{eqnarray}
 The $S^k_\pm(\mathbf{p})$
 have the same forms in equation (\ref{eq:pm})
 where the momentum operator $P^\mu$ is replaced by the
 momentum value $p^\mu$.
 We will use these representations in the specific frame for the case that
 the only spin context is needed, and also call $\Psi$ spin state.
 Under parity (spatial inversion), since the momentum and the PL vectors
 transform as $(P^0, \mathbf{P}) \leftrightarrow (P^0, -\mathbf{P})$ and
 $(W^0, \mathbf{W}) \leftrightarrow  (-W^0, \mathbf{W})$, respectively, and
 the spin operator $S^k_+$ transforms to the $S^k_-$ and vice versa in equation (\ref{eq:pm}).
 As the base spin states of the two inequivalent representations,
 the eigenstate $\Psi_+(p^\mu,\lambda_+)$
 then transforms to the $\Psi_-(p^\mu,\lambda_-)$ and vice versa
 for the same spin eigenvalue $\lambda$, i.e.,
 $\Psi_+(p^\mu,\lambda) \leftrightarrow \Psi_-(p^\mu,\lambda)$, under parity.
 In order to understand clearer how the two spin states
 $\Psi_+(p^\mu,\lambda_+)$ and  $\Psi_-(p^\mu,\lambda_-)$
 are related each other,
 we study the representations and their relation to
 the spin operators $S^k_\pm(\mathbf{p})$.

 At the PRF ($\mathbf{p}=0$) with the four-momentum $k^\mu=(m,
 \mathbf{0})$,
 since $S^k_+ (\mathbf{0})=S^k_- (\mathbf{0}) = W^k/m$,
 the eignestates becomes
 $\Psi_+ (k^\mu,\lambda) = \Psi_- (k^\mu,\lambda)\ (\equiv \Psi (k^\mu,\lambda))$
 for $\lambda_+ = \lambda_- (\equiv \lambda)$.
 Also $S^k_\pm(\mathbf{0})$ become a rotational generator $J^k$
 around $k$-th spatial coordinate because $W^k = m J^k$ at the PRF.
 The rotational generator $J^k$ at the PRF
 can be represented as the usual $\mathfrak{su}(2)$ operator $\sigma^k/2$,
 where $\sigma^k$ is the usual Pauli matrix satisfying
 the $\mathfrak{su}(2)$ algebra.
 Hence we naturally present the spin operator at PRF
 as the usual $\mathfrak{su}(2)$ operator $S^k_{\pm}(\mathbf{0})=\sigma^k/2$
 corresponding to the spin operator in the nonrelativistic quantum mechanics.
 In sequence, our purpose is to obtain the
 representation of the spin operators $S^k_\pm(\mathbf{p})$
 in an arbitrary moving frame. It can be accomplished most easily by
 using a Lorentz transformation (LT)
 of $p^\mu$ and $w^\mu$ in the representation space,
 changing over from the PRF to the moving frame with
 the momentum $\mathbf{p}$.
 Thus, we can consider only a boost transformation because
 a rotation in the PRF does not make the particle's momentum changed.
 Two successive non-collinear Lorentz boosts,
 equivalent to an effective rotation followed by an effective-single Lorentz boost,
 are well-known to give rise to a nontrivial effect \cite{Weinberg64}.
 However, such an effective rotation in the PRF is also not relevant to obtain
 the spin operators in the moving frame.
 Thus, the explicit expressions of two spin operators in the moving frame
 can be determined by a single pure boost transformation
 (so-called standard LT) $L(\mathbf{p})$, i.e.,
 $L^0_{\ 0} = p^0/m$, $L^0_{\ i} = p^i/m$, and $L^i_{j} = \delta_{ij} + p^i p^j / (m(p^0+m))$
 with the Kronecker-delta $\delta_{ij}$.
 This standard LT changes
 the particle momentum from $k^\mu=(m, \mathbf{0})$ to $p^\mu = (p^0, \mathbf{p})$
 as $p^\mu = L(\mathbf{p})^\mu_{\ \nu} k^\nu$ .
 At the PRF,
 the PL vector becomes $w^\mu_{rest} = (0, m \mbox{\boldmath $\sigma$}/2)$,
 because $w^0_{rest}=\frac{1}{2}\epsilon^{0ijk}J_{jk}k^i=0$ and
 $S^k_\pm(\mathbf{0}) = w^k_{rest}/m = J^k$ is represented by $\sigma^k /2$.
 The PL vector in the moving frame, then transformed by the
 standard LT, is given as
 $w^0 = L^0_{\ \mu} w^\mu_{rest} = ({\mbox{\boldmath $\sigma$} \cdot \mathbf{p}})/2$
 and $w^i =L^i_{\ \mu} w^\mu_{rest}
 = m \sigma^i/2 + p^i (\mbox{\boldmath $\sigma$} \cdot \mathbf{p})/(2(m+p^0))$.
 For the reference frame with the momentum $\mathbf{p}$,
 the spin operators in equation (\ref{eq:pm}) are represented as
  \begin{equation}
  S^k_\pm (\mathbf{p})
   = \frac{p^0}{2m} \sigma^k
        - \frac{ p^k (\mbox{\boldmath $\sigma$} \cdot \mathbf{p})} {2m (p^0 + m)}\
         {\pm}\ i \frac{1}{2m} \left(\mbox{\boldmath $\sigma$} \times \mathbf{p}\right)^k .
  \label{eq:spin}
  \end{equation}
 Note that
 these explicit
 representations of the $S^k_\pm(\mathbf{p})$
 provide the two inequivalent representations of the Poincar\'e group
 through the eigenvalue equations in equation (\ref{eq:sk}).
 The relation between the spin states $\Psi_+(p^\mu,\lambda)$ and $\Psi_-(p^\mu,\lambda)$
 for $s_+ = s_-$ in equation (\ref{eq:sk})
 can be understood by studying $(2s+1)$-dimensional transformation operators $ U_\pm[L(\mathbf{p})]$
 in obtaining
 $S^k_\pm (\mathbf{p}) $ from $S^k_\pm(\mathbf{0})$ in equation (\ref{eq:spin}), i.e.,
\begin{equation}
  S^k_\pm (\mathbf{p}) = U_\pm[L(\mathbf{p})] S^k_\pm(\mathbf{0}) U_\pm^{-1}[L(\mathbf{p})].
 \label{eq:covariance}
\end{equation}
 Actually,
 the explicit forms of the transformation operators $ U_\pm[L(\mathbf{p})]$
 are respectively derived (Methods) as
\begin{equation}
  U_\pm[L(\mathbf{p})] = \exp\left[{\pm\frac{1}{2}\boldsymbol{\sigma}\cdot \boldsymbol{\xi}}\right],
 \label{eq:LT}
\end{equation}
 where
 $\mbox{\boldmath $\xi$}= 2\, \hat{ \mathbf{p}} \tanh^{-1} [ |\mathbf{p}|/ (p^0+m) ]$.
 Then, the eigenstates $\Psi_\pm(p^\mu,\lambda)$
 in the moving frame and $\Psi(k^\mu,\lambda)$ at the PRF have the relations:
\begin{equation}
  \Psi_\pm(p^\mu,\lambda) = U_\pm[L(\mathbf{p})] \Psi(k^\mu,\lambda).
  \label{eq:TSPIN}
\end{equation}
 Hence, equation (\ref{eq:covariance}) ensures
 that the two eigenstates $\Psi_\pm(p^\mu,\lambda)$
 are respectively transformed from the eigenstate $\Psi(k^\mu,\lambda)$
 at the PRF without changing the spin eigenvalue $\lambda$.
 In this sense
 the transformation operators $U_\pm[L(\mathbf{p})]$ can be regarded
 as the spin state representation of the standard LT $L(\mathbf{p})$.
 Equation (\ref{eq:covariance}) also shows that
 the spin operators are related as $S^k_+ (-\mathbf{p}) = S^k_-(\mathbf{p})$
 because $U_+[L(-\mathbf{p})]= U_-[L(\mathbf{p})]$.
 More significantly,
 the transformation operators
 $U_\pm[L(\mathbf{p})]$ are the same
 as the right-handed and the left-handed representations of the standard LT
 $L(\mathbf{p})$, respectively,
 in the (homogeneous) Lorentz group \cite{Schwartz,Weinberg64}.
 Then the representation spaces whose base vectors are
 $\Psi_\pm(p^\mu,\lambda_\pm)$ respectively provide
 the right-handed and the left-handed representations
 in the Poincar\'e group.
 In contrast with that the Lorentz group is equivalent to $\mathrm{SU}(2)\times \mathrm{SU}(2)$
 and then its representation is the tensor product of
 the left-handed and the right-handed representations, it should be noted that
 the representation of the Poincar\'e group is given either by
 the right-handed or the left-handed representations
 because the second Casimir invariant of the Poincar\'e group is only one not two.
 Consequently,
 the space-time symmetry gives
 the two spin operators $S^k_\pm$ from the two physical requirements on the linear combination of PL vectors,
 the two spin operators determine the two inequivalent representations
 of the Poincar\'e group,
 and the two inequivalent representations are
 identified by the handedness of the spin state
 for arbitrary spin massive particles.
 The two representation spaces of the Poincar\'e group play a fundamental role
 as the building blocks for the irreducible representations.

 Under parity, as we discussed,
 the base states $\Psi_+(p^\mu,\lambda)$ and $\Psi_-(p^\mu,\lambda)$
 are interchanged each other.
 In describing a free massive elementary particle with parity symmetry,
 the representation space should then contain
 all states of the two types of $\Psi_+(p^\mu,\lambda)$ and $\Psi_-(p^\mu,\lambda)$.
 This requires that the Poincar\'e group is extended by parity,
 and the natural representations for the parity-extended Poincar\'e group
 are the tensor product
 of the left-handed and the right-handed representations, i.e., $(s_-,s_+)$ representations
 similar to the representations of the Lorentz group.
 However, the symmetry under the parity operation
 does not allow all the tensor product representations.
 In conclusion, for a complete description of free massive elementary particles,
 the Poincar\'e group is extended by parity, and
 the possible natural representations of the parity-extended Poincar\'e group
 are non-chiral $(s,s)$ representations
 and direct-sum $(s_-,s_+)\oplus(s_+,s_-)$ representations.
 This makes the reason clear, for instance, why massive elementary particles with spin-$1/2$,
 i.e., the Dirac particles,
 are well described in the direct-sum $(1/2,0)\oplus(0,1/2)$ representation
 as one of the natural representations of the parity-extended Poincar\'e group.

\noindent\textbf{Little groups generated by the two spin operators.}
 Satisfying the $\mathfrak{su}(2)$ algebra,
 each of $S^k_\pm$ generates a $\mathrm{SU}(2)$ group.
 The elements of this $\mathrm{SU}(2)$ groups can be denoted by
 $\mathcal{D}_\pm(\theta^k_\pm) = \exp\left[\frac{i}{2} \theta^k_\pm S^k_\pm \right]$
 with a finite parameter $\theta^k_\pm$ of the rotation group.
 The group elements $\mathcal{D}_\pm(\theta^k_\pm)$
 do not change the momentum of a particle
 because
\begin{eqnarray}
 \mathcal{D}_\pm(\theta^k_\pm)  P^\mu \mathcal{D}^{-1}_\pm(\theta^k_\pm) = P^\mu,
\end{eqnarray}
 which is guaranteed by $[S^k_\pm, P^\mu]=0$.
 As is known, the subgroup of the Lorentz group
 that does not change the momentum of a particle is called the little
 group \cite{Wigner}.
 To complete our argument that
 the group composed of every element $\mathcal{D}_\pm(\theta^k_\pm)$ is a little group,
 we have shown that the action of $\mathcal{D}_\pm(\theta^k_\pm)$ on the spin states
 is represented by LTs (Methods).

 In general, the base sates
 in both the right-handed and the left-handed representations
 undergo the same little group rotation for general LTs (Methods).
 Due to the little group rotation of the spin states,
 the consideration of a spin-state projected Lagrangian
 is not physically meaningful in relativistic situation \cite{Choi14}.
 The fact that the rotation angles $\theta^k_\pm$ of
 the little groups corresponding to a specific LT
 are the same both for the right-handed and the left-handed states,
 seems to be consistent with the existence of
 only one little group in Wigner's representation \cite{Wigner}.
 Actually,
  the representation of our little group differs from
 that of Wigner little group
 because
 the representation of our little group is not unitary
 in contrast to the representation of Wigner little group.
 Both of the two little groups generated by $S^k_\pm$ are $\mathrm{SU}(2)$ groups.
 One can notice that the spin operators $S^k_\pm({\bf p})$ are non-Hermitian and
 the base states $\Psi_\pm(p^\mu,\lambda)$ are not unitary.
 This fact is very important to understand
 why the Dirac spinor as the solution of the Dirac equation is non-unitary.
 It can be verified through showing that the Dirac equation
 is equivalent to the fundamental equation obtained
 from the covariant parity equation given
 by our spin operators in the next discussion.

%


\noindent\textbf{ The spin operators and the covariant fundamental
 equation equivalent to the Dirac equation.}
 Among our possible natural representations of the parity-extended Poincar\'e group
 for a complete description of free massive particles with spin $s$,
 the direct sum $(s,0) \oplus (0,s)$ representation is
 the only representation without any redundant representation space.
 Spin-$s$ free massive elementary particles
 can then be rightly described in our direct sum $(s,0) \oplus (0,s)$
 representation.
 The spin operator in terms of $S^k_\pm$ in the $2(2s+1)$-dimensional representation
 is given as
\begin{eqnarray}
\label{eq:DSP}
 \mathcal{S}^k =S^k_-\oplus S^k_+ = \frac{\mathbb{I}_{2(2s+1)}}{m^2}(P^0 W^k -P^k W^0)
  + i \frac{\gamma^5}{m^2}\epsilon_{0kml}W^m P^l,
\end{eqnarray}
 with the $2(2s+1)$-dimensional identity matrix $\mathbb{I}_{2(2s+1)}$
 and
 $\gamma^5= \left( \begin{array}{cc}  -\mathbb{I}_{2s+1} & 0 \\ 0 & \mathbb{I}_{2s+1} \end{array} \right)$.
 One can show clearly
 that the spin three vector $\mathcal{S}^k$ in Eq. (\ref{eq:DSP})
 satisfies both the $\mathfrak{su}(2)$ algebra
 and the Lorentz-transformation properties as the tensorial requirement.
 Obviously, the Casimir operators are $P^\mu P_\mu$ and
 $\mathcal{S}^k \mathcal{S}^k$
 in the parity-extended Poincar\'e group.
 The direct sum $(s,0) \oplus (0,s)$ representation
 is the irreducible representation of the parity-extended Poincar\'e group
 for a massive particle with the mass $m$ and the spin $s$,
 which is labelled by the eigenvalues of $\{ P^\mu, \mathcal{S}^k\}$.
 The eigenvalue equations for the $(s,0)\oplus (0,s)$ representation can be written as
\begin{subequations}
\begin{eqnarray}
 P^\mu \ \psi_D(p^\mu,\lambda) &=& p^\mu\ \psi_D(p^\mu,\lambda),
 \label{eq:p_direct}
 \\
 \mathcal{S}^k (\mathbf{p}) \psi_D(p^\mu,\lambda) &=& \lambda\ \ \psi_D(p^\mu,\lambda),
 \label{eq:sk_direct}
\end{eqnarray}
 where with $\lambda \in \{ -s, -s+1, \cdots, s\}$ both for the left-handed and the right-handed spin states,
 the space of base states in
 the $(s,0) \oplus (0,s) $ representation space are composed
 of a linear combination of the direct-sum states of
 the eigenstates in equation (\ref{eq:sk}):
\begin{eqnarray}
 \psi_D(p^\mu,\lambda)
 = \left( \begin{array}{c} \Psi_-(p^\mu,\lambda) \\ \Psi_+(p^\mu,\lambda) \end{array} \right)
 =  U [L(\mathbf{p})] \psi_D(k^\mu,\lambda).
 \label{eq:dstates}
\end{eqnarray}
\end{subequations}
 Here, $\psi_D(k^\mu,\lambda)$ is the eigenstate at PRF and the $2(2s+1)$-dimensional standard LT is given in
 the direct-sum representation as $U[L(\mathbf{p})] = U_-[L(\mathbf{p})] \oplus U_+[L(\mathbf{p})]$.
 Since the $\mathcal{S}^k$ generates a $\mathrm{SU}(2)$ subgroup in the parity-extended Poincar\'e group,
 whose elements can be denoted by $\mathcal{D} (\theta^k) = \exp[\frac{i}{2} \theta^k  \mathcal{S}^k]$
 with the angle $\theta^k$, the group composed of every element $\mathcal{D}(\theta^k)$
 is a little group of the parity-extended Poincar\'e group
 because $\mathcal{D} P^\mu \mathcal{D}^{-1} = P^\mu$ due to $[\mathcal{S}^k, P^\mu] =0$.

 An equation of motion for free elementary particles,
 such as the Klein-Gordon equation
 and the original Dirac equation given by the relativistic
 invariant relation of energy-momentum \cite{Schwartz}, is expected to be
 derived from the spacetime symmetry relations.
 Note that since $\Psi_\pm(\tilde p^\mu,\lambda) = \Psi_\mp (p^\mu,\lambda)$ with
 $\tilde p^\mu=(p^0, -{\bf p})$,
 as usual, one may regard the parity operation ${\mathcal P}$ as
\begin{eqnarray}
 \mathcal{P}\, \psi_D(p^\mu, \lambda)
  = \left( \begin{array}{c} \Psi_+(p^\mu,\lambda) \\
                             \Psi_-(p^\mu,\lambda) \end{array}
  \right)
 \rightarrow \gamma^0 \psi_D(p^\mu,\lambda),
 \label{eq:PTS}
\end{eqnarray}
 where
 $\gamma^0 = \left( \begin{array}{cc} 0 & \mathbb{I}_{2s+1} \\
 \mathbb{I}_{2s+1} & 0 \end{array} \right)$.
 Equation (\ref{eq:PTS}) is an another nontrivial relation given by
 parity operation for free massive particles with spin.
 However, the $\gamma^0$ is not a covariant representation of the parity operation
 $\mathcal{P}$.
 Since not only parity but also Lorentz symmetries are included
 in the extended Poincar\'e symmetry,
 the representation of parity operator $\mathcal{P}$
 should be covariant under LT.
 Because the representation space is constructed by the eigenstates of $P^\mu$
 and $S^k_\pm$, the covariant form of the parity operator in
 the spin state space should be also constructed by $P^\mu(p^\mu)$ and
 $S^k_\pm (S^k_\pm(\mathbf{p}))$.
 Owing to
 the transformation properties of the spin states $\Psi_\pm(p^\mu,\lambda)$
 under $U_\pm[L(\mathbf{p})]$,
 the covariant parity operation on the spin eigenstates can be obtained.
 However, only for spin-$1/2$ case,
 the covariant parity operation is possible and defined (Methods) as
\begin{equation}
 \frac{1}{m} \Big( p^0 + 2 \mathfrak{S}_{0\mu}({\bf p})p^\mu \Big)
 \Psi_\pm (p^\mu,\lambda) = \Psi_\mp (p^\mu,\lambda),
\end{equation}
 where
 the antisymmetric tensor operator $\mathfrak{S}_{\nu\mu}$
 is defined by using the dual spin tensor operator $^*S_{\nu\mu;\pm}$ as
\begin{equation}
\mathfrak{S}_{\nu\mu}({\bf p})\Psi_\pm(p^\mu,\lambda)=\mp
^*S_{\nu\mu;\pm}({\bf p}) \Psi_\pm(p^\mu,\lambda).
\end{equation}
 Here
 $S^k_\pm = \frac{1}{2} \epsilon_{klm} S_{lm;\pm}$
 with the spin tensor $S^{\mu\nu}_\pm$
 and  the dual spin tensor
 $ {}^*S^{\mu\nu;\pm}=\frac{1}{2}\epsilon^{\mu\nu\rho\sigma} S_{\rho\sigma;\pm}$,
 and then
 $\mathfrak{S}_{0k}({\bf p})\Psi_\pm(p^\mu,\lambda)
 =\mp S^k_\pm({\bf p}) \Psi_\pm(p^\mu,\lambda)$.
 Hence only for spin-$1/2$ massive particles,
 the covariant form and operation of the parity operator in
 the $(1/2,0)\oplus(0,1/2)$ representation are represented as
\begin{equation}
 \mathcal{P}\, \psi_D(p^\mu,\lambda)
 = \frac{1}{m}\!\left[ p^0 + 2\!\left( \begin{array}{cc}  \mathfrak{S}_{0\mu}({\bf p}) p^\mu& 0 \\ 0
& \mathfrak{S}_{0\mu}({\bf p})p^\mu \end{array} \right)\right] \! \psi_D(\lambda;p)
 = \gamma^0 \psi_D(p^\mu,\lambda).
 \label{eq:CPAO}
 \end{equation}
 This covariant representation of the parity operator transforms
 as $0$-th component of a four-vector under a LT.
 To make a compact form of equation (\ref{eq:CPAO}),
 we multiply both sides of equation (\ref{eq:CPAO}) by $\gamma^0$
 and then obtain the Lorentz covariant equation
\begin{subequations}
\begin{equation}
\label{eq:FOESH}
 \left( \tilde{\gamma}^\mu(\mathbf{p})\; p_\mu -m \right) \psi_D(p^\mu,\lambda)=0,
\end{equation}
 where the defined $\tilde{\gamma}^\mu$ matrices are given by
\begin{eqnarray}
 \tilde{\gamma}^0 &=& \gamma^0 \mbox{~and~}
 \tilde{\gamma}^k (\mathbf{p})
 = 2\left( \begin{array}{cc} 0 & \mathfrak{S}^{0k}({\bf p})
     \\ \mathfrak{S}^{0k}({\bf p}) & 0 \end{array} \right).
\label{eq:gamma}
\end{eqnarray}
\end{subequations}
 Consequently, we obtain the fundamental equation for a free massive particle
 with spin $1/2$ in equations (\ref{eq:FOESH}) and (\ref{eq:gamma})
 from the covariant parity operation on the direct sum
 $(1/2,0) \oplus (0,1/2)$ representation.
 One can confirm easily that the parity-inversion spin state
 $\psi^\mathcal{P}_D(p^\mu,\lambda)=\gamma^0 \psi_D(p^\mu,\lambda)$
 satisfies the same equation in equations (\ref{eq:FOESH}) and (\ref{eq:gamma}),
 i.e.,
 $(\tilde{\gamma}^\mu (\mathbf{p})\, p_\mu -m) \psi^\mathcal{P}_D(p^\mu,\lambda)=0$.
 It should be noted that the new gamma matrices also satisfy
 the Clifford algebra, i.e.,
 $\tilde \gamma^\mu (\mathbf{p}) \tilde \gamma^\nu (\mathbf{p})
  + \tilde \gamma^\nu (\mathbf{p}) \tilde \gamma^\mu (\mathbf{p}) = g^{\mu\nu}$.
 The fundamental equation shows that
 the relativistic spin $\mathfrak{S}_{0\mu}$ is naturally coupled
 with the momentum $p^\mu$
 for free spin-$1/2$ massive particles.

 The fundamental equation in equations (\ref{eq:FOESH}) and (\ref{eq:gamma}) for spin $1/2$
 seems to be higher-order equation rather than a first-order equation
 in the momentum
 because the tensor operator $\mathfrak{S}^{0k}(\mathbf{p})$
 depends on the momentum $p^\mu$ through the spin operators
 $S^k_\pm(\mathbf{p})$.
 Very interestingly, however,
 one may notice the equivalence $2 S^k_\pm({\bf p}) p^k=\sigma^k p^k$ for any spin $s$.
 In the fundamental equation,
 our gamma matrices $\tilde \gamma (\mathbf{p})$ in equation (\ref{eq:gamma}) can
 then be reduced to the usual Dirac gamma matrices $\gamma^\mu$
 described by only the Pauli matrices.
 Hence, our fundamental equation in equations (\ref{eq:FOESH}) and (\ref{eq:gamma})
 for a free massive particle with spin $1/2$ is equivalent to
 the usual covariant Dirac equation
 $\left( \gamma^\mu p_\mu - m \right)\psi =0$  \cite{Schwartz}.
 It is shown that the covariant Dirac equation
 is naturally given by the space-time symmetry, i.e.,
 the Poincar\'e symmetry extended by parity symmetry,
 and  the appearance of the Pauli matrices \cite{Schwartz} in it
 is a natural consequence of the relativistic spin operators
 $S^k_\pm(\mathbf{p})$.
 This fact explains clearly why the Dirac equation can predict the existence of spin
 and it describes spin-$1/2$ massive elementary particles very well.
 Additionally,
 the noticeable property of the relativistic spin $\mathbf{S}_\pm(\mathbf{p})$ for any spin $s$
 is to be the equivalence of its projection
 to the projection of known spin matrices
 onto the particle's spatial momentum,
 i.e., the helicity operator
 $\boldsymbol{\sigma}/2 \cdot \mathbf{p}/|\mathbf{p}| $ \cite{Schwartz}.
 This fact makes the use of the helicity operator justified for
 relativistic massive particles with any spin $s$.

\noindent\textbf{Proper position operator and nonexistence of
 Zitterbewegung.}
 So far we have presented the rigorous derivation of spin operators,
 the corresponding natural representations of Poncar\'e group,
 the origin of particle's handedness, and the fundamental equation
 from the covariant parity operation.
 These give clear answers on the fundamental questions associated
 with spin.
 Especially,
 the newly derived fundamental equation in equations (\ref{eq:FOESH}) and (\ref{eq:gamma})
 manifests
 the clear-cut verifications of the origin of spin
 and provide more profound understandings
 on the physical results expected by the original Dirac theory.
 Further, our relativistic spin can provide new physical understandings
 on the emerging inconsistencies and the controversial issues.
 Of significant topic for an origin of spin for $s=1/2$ is the Zitterbewegung
 predicted by Schr\"odinger \cite{Schrodinger}.
 To be interpreted as an origin of spin for $s=1/2$, i.e., an intrinsic nature of elementary particles,
 the Zitterbewegung is however problematic
 because it survives
 only if there occurs an interference between the positive-
 and negative-energy eigenstates of the solutions of the Dirac equation \cite{Thaller}.
 More paradoxically,
 the acceleration of the free particle is not zero \cite{Thaller}.
 The standard position operator $\mathbf{X}$, being used in
 nonrelativistic quantum theory, has been used to actually result
 in the prediction conflicting severely with the Newton's
 second law of motion \cite{Thaller}.
 The Zitterbewegung is essentially related to the fundamental problem
 of proper position operator in relativistic quantum theory.
 Since the (proper) spin operator is discovered for massive particles,
 the total angular momentum as the constant of motion in the Dirac Hamiltonian
 can offer a way to determine the corresponding proper relativistic position operator
 that resolves
 the paradoxical relativistic quivering motion of free Dirac
 particles
 described by using the improper position operator.

 Dirac \cite{Dirac} found that
 the total angular momentum
 $\mathbf{J} = \boldsymbol{\Sigma}/2 + \mathbf{X} \times \mathbf{P}$
 is a constant of motion, i.e., it
 commutes with the Dirac Hamiltonian
 $H = \boldsymbol{\alpha}\cdot \mathbf{P}+\beta\, m$,
 i.e., $\left[ H, \mathbf{J} \right] = 0$,
 where $ {\boldsymbol{\alpha}}=\gamma^0\boldsymbol{\gamma}$,
 $\beta=\gamma^0$, and $\boldsymbol{\Sigma}=\gamma^5 {\boldsymbol{\alpha}}$.
 Using the relativistic spin operator $\boldsymbol{\cal S}$ in equation (\ref{eq:DSP}),
 the total angular momentum $\mathbf{J}$ can be decomposed
 as $\mathbf{J}= \boldsymbol{\cal S} + \boldsymbol{\cal X} \times \mathbf{P}$
 with  a proper relativistic position operator $\boldsymbol{\cal
 X}$.
 Similar to the standard position operator $\mathbf{X}$,
 the proper position operator $\boldsymbol{\cal X}$
 should also satisfy  $[{\cal X}^i,{\cal X}^j]=0$ for the locality requirement \cite{NW}.
 Satisfying the requirements, the proper position operator is obtained as
\begin{eqnarray}
 \boldsymbol{\cal X}=\mathbf{X} + \frac{\boldsymbol{\Sigma}\times
 {\bf P}}{2m(P^0+m)} - i\frac{\gamma^5\boldsymbol{\Sigma}}{2m}
 +i \frac{ \gamma^5 (\boldsymbol{\Sigma}\cdot{\bf P}){\bf P}}{2m P^0(P^0+m)}.
\end{eqnarray}
 This relativistic position operator satisfies the relation
 $\boldsymbol{\cal X} =
 \exp\left[\gamma^5 \boldsymbol{\Sigma}\cdot\boldsymbol{\xi}/2\right]
 \, \mathbf{X}
 \exp\left[-\gamma^5 \boldsymbol{\Sigma}\cdot\boldsymbol{\xi}/2\right]$,
 where ${\cal U} =
 \exp\left[\gamma^5 \boldsymbol{\Sigma}\cdot\boldsymbol{\xi}/2\right]$
 is the spin state representation of the Lorentz boost
 with the rapidity $\boldsymbol{\xi}$
 in the direct sum $(1/2,0)\oplus(0,1/2)$ representation.
 This seems to imply that the relativistic position operator $\boldsymbol{\cal X}$
 in the momentum representation can be considered as the position operator
 Lorentz-boosted from the position operator $\mathbf{X}$ at the rest frame,
 similar to that the spin operator $\boldsymbol{\cal S}$ can be regarded as
 the spin operator in the moving frame boosted from the spin operator
 $\boldsymbol{\Sigma}$ at the rest frame.
 However, the standard position operator $\mathbf{X}$ is not the position operator
 at the rest frame but the position operator conjugates to the momentum, i.e.,
 $i\boldsymbol{\partial_p}$ in the momentum representation.

 Straightforwardly, the velocity operator
 $\boldsymbol{\upsilon} = d\boldsymbol{\cal X}/dt = i[H, \boldsymbol{\cal X}]$
 for free Dirac particles is calculated by using
 the commutation relation with the free Dirac Hamiltonian $H$ as
\begin{eqnarray}
 \frac{d\boldsymbol{\cal X}}{dt}
 =\frac{P^0}{m}\boldsymbol{\alpha} - \frac{ (\boldsymbol{\gamma}\cdot {\bf P}){\bf P}}{P^0(P^0+m)}
 +i \frac{\boldsymbol{\Sigma}\times {\bf P}}{m} + \boldsymbol{\gamma}
 - \frac{(\boldsymbol{\alpha}\cdot {\bf P}){\bf P} }{m(P^0+m)}.
\end{eqnarray}
 The eigenvalues of the velocity operator $\boldsymbol{\upsilon}$ are
 $\pm {\bf p}/p^0$ in a sharp contrast to the eigenvalues $\pm 1$
 of the $d\mathbf{X}/dt=\boldsymbol{\alpha}$ expected by Schr\"odinger.
 The eigenvalues of the velocity operator $\boldsymbol{\upsilon}$ depend only
 on the momentum so that the expectation value of the commutator
 between the velocity operator and the
 free Dirac Hamiltonian is zero, i.e.,
 the velocity operator $\boldsymbol{\upsilon}$ is a constant of motion.
 This can be confirmed by calculating the acceleration operator $d\boldsymbol{\upsilon}/dt = i\left[H, d\boldsymbol{\cal X}/dt\right]$:
\begin{eqnarray}
\frac{d^2\boldsymbol{\cal X}}{dt^2}
 =\!  \frac{2P^0}{m} {\bf P}\! \times\! \boldsymbol{\Sigma}
 +\! 2i\! \left[ \frac{(\mathbf{P}\! \times\! \boldsymbol{\gamma})\! \times\! \mathbf{P}}{P^0+m}
 -\!\frac{m\beta \mathbf{P}}{P^0} -\! \frac{m(\boldsymbol{\alpha}\cdot\mathbf{P})\mathbf{P}}{P^0(P^0+m)}
 +\!\frac{(\mathbf{P}\!\times\!\boldsymbol{\alpha})\!\times\!\mathbf{P}}{m}
 +\! m(\beta\!+\! \mathbb{I}_4)\boldsymbol{\gamma}\! \right]\! .
\end{eqnarray}
 Compared to the non-zero eigenvalue
 of the $d^2\mathbf{X}/dt^2$,
 note that the eigenvalues of the acceleration operator
 $d\boldsymbol{\upsilon}/dt$ are zero,
 which means that the expectation value of the acceleration operator
 on free Dirac particle states
 is zero.
 Consequently,
 the expectation value of the velocity operator
 for free Dirac particles is constant
 with the value $\mathbf{p}/p^0$ as the classical velocity.
 With the constant momentum $d\mathbf{P}/dt = i[H,\mathbf{P}]=0$,
 the physical motion of free Dirac particles
 does not conflict with the Newton's first
 law of motion.
 We have shown that no quivering motion (Zitterbewegung)
 appears at all
 for free Dirac particles as a representation of the extended Poincar\'e group
 of a four-dimensional space-time symmetry,
 even for the case with both particle and antiparticle,
 which are represented by four-spinors.
 The Zitterbewegung appearing through the Heisenberg equation of the standard position operator $\mathbf{X}$
 can be concluded not to be a physical motion for a free Dirac particle.

\noindent\textbf{Conserved spin current from Noether's theorem.}
 Of important issue is whether the spin as an intrinsic kinematic
 property of massive particles is a conserved quantity
 because the associated conservation laws,
 as fundamental features of our four-dimensional spacetime,
 can play a central role in the relativistic quantum theory.
 In general, Noether's method \cite{Noether} allows us to answer explicitly on this question
 with considering the spin-$1/2$ Lagrangian
 ${\cal L} = \bar\psi_D (i\tilde \gamma^\mu \partial_\mu-m)\psi_D$
 that gives the fundamental equation in equations (\ref{eq:FOESH}) and (\ref{eq:gamma}),
 where $\bar\psi_D = \psi^\dagger \gamma^0$.
 Because of $\tilde \gamma^\mu p_\mu = \gamma^\mu p_\mu$, the
 Lagrangian is equal to the usual QED Lagrangian
 ${\cal L} = \bar\psi_D (i\gamma^\mu \partial_\mu-m)\psi_D$.
 For Lorentz transformation in the Poincar\'e group,
 Noether's theorem \cite{Noether} gives
 the conserved current
\begin{equation}
 ({\cal J}^\mu)^{\rho\sigma}
 = X^\rho T^{\mu\sigma} - X^\sigma T^{\mu\rho}
  +\bar\psi_D \gamma^\mu \Sigma^{\rho\sigma}\psi_D,
  \label{eq:Totalcurrent}
\end{equation}
 which satisfies $\partial_\mu ({\cal J}^\mu)^{\rho\sigma} =0$,
 where the energy-momentum tensor
 $T^{\mu\nu} = i\bar\psi_D  \gamma^\mu \partial^\nu \psi_D$
 and the Lorentz generator $\Sigma^{\rho\sigma} = \frac{i}{4}\left[ \gamma^\rho,
 \gamma^\sigma\right]$.
 The conserved current gives rise to the conserved charges that
 consist of the total angular momentum
 $Q^{ij} = \int d^3x ({\cal J}^0)^{ij}= \int d^3x \, (X^i T^{0j} - X^j T^{0 i}
  +\bar\psi_D \gamma^0 \Sigma^{ij}\psi_D)$
  and a conserved quantity under pure boosts
  $Q^{0i} = \int d^3x ({\cal J}^0)^{0i}= \int d^3x \, (X^0 T^{0i} - X^i T^{0 0}
  +\bar\psi_D \gamma^0 \Sigma^{0i}\psi_D)$.
 One can confirm that the last term of equation
 (\ref{eq:Totalcurrent}) comes from the Lorentz transformation of
 the spin states and
 does not satisfy itself
 the current conservation, i.e., $\partial_\mu (i\bar\psi_D \gamma^\mu
 \Sigma^{\rho\sigma}\psi_D)=0$.
 To make the contribution of spin clear for the Noether's current
 under the Lorentz symmetry,
 thus let us decompose the conserved current of equation
 (\ref{eq:Totalcurrent}) properly by using the spin
 operator as follows:
\begin{equation}
 ({\cal J}^\mu)^{\rho\sigma}
 = {\cal X}^\rho T^{\mu\sigma} - {\cal X}^\sigma T^{\mu\rho}
  +\bar\psi_D \gamma^\mu {\cal S}^{\rho\sigma}\psi_D,
 \label{eq:Totalcurrent2}
\end{equation}
 where ${\cal X}^i$ is a corresponding proper position
 operator.
 Then the current for a proper contribution of spin can
 be defined by
\begin{equation}
 \label{eq:Scurrent}
 \left( {\cal J}^{\mu}_S\right)^{\rho\sigma}
  = \bar\psi_D \gamma^\mu {\cal S}^{\rho\sigma} \psi_D ,
\end{equation}
 where the spin tensor
 ${\cal S}^{\rho\sigma} = \frac{i}{4}\left[ \tilde \gamma^\rho,
 \tilde \gamma^\sigma\right]$.
 To be conserved itself, the current in equation (\ref{eq:Scurrent}) should satisfy
 $\partial_\mu ( {\cal J}^\mu_S )^{\rho\sigma} =0$, i.e.,
 $ \bar\psi_D[ \mathcal{S}^{\rho\sigma} , \gamma^{\mu} P_{\mu} ]\psi_D=0$.
 The requirement can be rewritten as
 $\bar \psi_D [ \mathcal{S}^{\rho\sigma} , \gamma^{\mu} P_{\mu}] \psi_D=
 \bar \psi_D [ \mathcal{S}^{\rho\sigma}_{rest}, m \gamma^0 ]
 \psi_D$ by considering
 a LT $U[L(\mathbf{p})]=U_-[L(\mathbf{p})]\oplus U_+[L(\mathbf{p})]$
 transforming from the PRF
 to an arbitrary reference frame
 and using the relations
 $\mathcal{S}^{\rho\sigma}
 =U[L(\mathbf{p})] \mathcal{S}^{\rho\sigma}_{rest} U^{-1}[L(\mathbf{p})]$,
 $U[L(\mathbf{p})] P_\mu U^{-1} [L(\mathbf{p})]= L_\mu^{\phantom{\mu} \rho} P_\rho$,
 and $U[L(\mathbf{p})] \gamma^{\mu} U^{-1}[L(\mathbf{p})]= L_\sigma^{\phantom{\sigma} \mu}
 \gamma^{\sigma}$,
 where ${\cal S}^{\rho\sigma}_{rest} = \Sigma^{\rho\sigma}/2$.
 We find that
  $\bar\psi_D[ \mathcal{S}^{0i}, \gamma^{\mu} p_{\mu} ]\psi_D \neq 0$
 because $\bar\psi_D[ \mathcal{S}^{0i}_{rest}, m \gamma^0 ]\psi_D \neq 0$.
 This implies that under pure boots,
 the $({\cal J}^\mu)^{0i}$ is only conserved itself.

 However, we find that
 $\bar\psi_D[ \mathcal{S}^{ij}, \gamma^{\mu} p_{\mu} ]\psi_D = 0$
 because $\bar\psi_D [ \mathcal{S}^{ij}_{rest}, m \gamma^0 ]\psi_D = 0$.
 The $({\cal J}^\mu)^{ij}$, giving rise to the total angular momentum
 as the conserved charge,
 are decomposed into the two conserved
 currents:
\begin{subequations}
 \begin{equation}
 ({\cal J}^\mu)^{ij} = ({\cal J}^\mu_L)^{ij} + ({\cal
 J}^\mu_S)^{ij},
 \end{equation}
 where the relativistic orbital current
 $({\cal J}^\mu_L)^{ij}$ and the relativistic spin current $({\cal J}^\mu_S)^{ij}$
 are given as
 \begin{eqnarray}
 ({\cal J}^\mu_L)^{ij} &=& {\cal X}^i T^{\mu\, j} - {\cal X}^j T^{\mu\, i},
 \label{eq:ScurrentL}
 \\
 ({\cal J}^\mu_S)^{ij} &=& \bar\psi_D \gamma^\mu {\cal S}^{ij}\psi_D,
 \label{eq:ScurrentS}
 \end{eqnarray}
 satisfying $\partial_\mu({\cal J}^\mu_L)^{ij} =0$ and $\partial_\mu ({\cal J}^\mu_S)^{ij}
 =0$,
 respectively.
\end{subequations}
 The relativistic orbital and spin currents
 give rise to the corresponding conserved charges, i.e.,
 the orbital angular momentum
 $Q^{ij}_L = \int d^3x ({\cal J}^0_L)^{ij}= \int d^3x({\cal X}^i T^{0\, j} - {\cal X}^j T^{0\, i})$
 and the spin angular momentum
 $Q^{ij}_S = \int d^3x ({\cal J}^0_S)^{ij} = \int d^3x\, \psi^\dagger_D {\cal S}^{ij}
 \psi_D$, respectively.
 Therefore,
 the relativistic spin $\mathcal{S}^k (=\frac{1}{2} \epsilon_{kij}{\cal S}^{ij})$ for spin $1/2$
 in equation (\ref{eq:DSP}) is a good observable
 and the relativistic spin current $(J^\mu_S)^{ij}$ in equation (\ref{eq:ScurrentS})
 is conserved.
 These are a natural consequence of the fact that
 the spin operators ${\cal S}^{k}$ are the generators of the little group
 as a subgroup of
 the parity-extended Poincar\'e symmetry group.
 It should be also noted that
 like the case of non-relativistic systems where one can specify a given
 energy state by the projection of spin along the $z$-axis (namely, by
 the eigenvalue of $\mathcal{S}_z$), in the relativistic case such
 a specification is useful since
 spin is a constant of motion.

\noindent\textbf{Discussion}

\noindent
 For discussing more implications of our results,
 it would be better to summarize our results in a view of consequentially developed fundamental concepts.
 We have derived the relativistic spin operators for any spin $s$ from the physical requirements on the spacetime symmetry,
 which shows that
 the intrinsic angular momentum of massive elementary particles
 are a relativistic-kinematic quantum character of spacetime.
 In contrast to a common belief, there are the two relativistic spin operators
 for massive particles with any integer or half-integer spin.
 The two spin operators are also the generators of which the little groups
 are a subgroup of Lorentz transformations which leave the momentum of a particle invariant.
 The handedness of massive particles arises naturally
 as a consequence of the two relativistic spin operators in the representation of the Poincar\'e group.
 Under the parity transformation, moreover, one handedness representation turns into the other handedness representation due to that one spin operator becomes
 the other spin operator.
 This fact implies that free massive particles with spin $s$ are completely described in the parity-extended Poincar\'e group
 and the only natural representation without any redundant representation space
 is the direct sum $(s,0)\oplus(0,s)$ representation.

 Furthermore, the covariant parity operation on the direct sum $(s,0)\oplus(0,s)$ representation
 can provide a fundamental equation for massive particles with any spin $s$,
 together with the eigenvalue equations (\ref{eq:p_direct}) and (\ref{eq:sk_direct}) with
 respect to the momentum $P^\mu$ and the spin angular momentum $\mathcal{S}^k$, respectively.
 However, only for spin-$1/2$ massive particles,
 the covariant parity operator, which is derived from the covariance requirements,
 exists and then results in the fundamental equation.
 We find that the fundamental equation for spin $1/2$
 is equivalent to the covariant Dirac equation
 originally derived as a first-order equation satisfying
 the Einstein's energy-momentum relation.
 In contrast to the existence of the Dirac equation for spin-$1/2$ massive particles,
 the nonexistence of covariant operators
 for massive particles with spin higher than $1/2$
 may explain why for higher spin massive particles, as is well-known,
 a proper relativistic description has still been a long standing problem in quantum theory.
 As an example,
 although our direct sum states $\psi_D$ satisfy the relativistic wave equations suggested
 by Weinberg\cite{Weinberg64} for spin-$s$ massive particles,
 the unphysical solutions \cite{Ahluwalia} have been noticed in the Weinberg equations
 due to $2s$-order time derivatives.
 Other most proposed descriptions have been known to have
 the main drawbacks such as containing redundant or unwanted degrees of freedom.
 However, exploring higher spin massive particles in a more systematic way
 has become possible with our established facts that for any spin $s$, the spin operators are given explicitly in equation (\ref{eq:pm}),
 they satisfy the eigenvalue equation in equation (\ref{eq:sk}),
 and their projection on the momentum is equal to the conventional definition of particle's helicity,
 i.e. $2 S^k_\pm({\bf p}) p^k=\sigma^k p^k$.

%
 As the paradoxical prediction from the Dirac equation for spin $1/2$,
 the Schr\"odinger's Zitterbewegung is resolved not to happen in accordance with the proper
 relativistic position operator
 which is determined by the relativistic spin operator through the total angular momentum for free Dirac particles.
 Whether the Zitterbewegung exists has recently been a revived interest in simulations of relativistic quantum effects
 using different artificial and nonrelativistic physical systems.
 Interestingly, a one-dimensional spinless Dirac-like Hamiltonian
 satisfying the one-dimensional energy-momentum relation
 has been realized in single ion trap experiments \cite{Gerritsma}
 and an oscillatory behavior similar to the original Zitterbewegung
 has been observed in the average value of the spinless particle's position
 due to a quantum interference between
 mimic positive- and negative-energy states, which depends on the
 setup of the initial state.
 Thus it is pertinent to mention $1+1$ dimensional spacetime symmetry
 for comparison to $3+1$ dimensional spacetime symmetry  for relativistic observables.
 For the Poincar\'e group of a $1+1$ dimensional spacetime symmetry,
 there is only one Casimir operator for mass, i.e., $P^\mu P_\mu$ with the eigenvalue $m^2$ with $\mu \in \{0,1\}$ \cite{Bekaert}.
 No spin exists for $1+1$ dimensional particles.
 Similar to the $3+1$ dimensional Dirac equation,
 $1+1$ dimensional Dirac-like equation has solutions of positive- and negative-energy states,
 although it does not have spin \cite{Thaller2}.
 Then, $1+1$ dimensional spinless Dirac particle was also expected to have a
 similar Zitterbewegung based on the one-dimensional standard position operator $X$
 with similar paradoxical features \cite{Thaller2}.
 The experimental result supports that
 a Zitterbewegung phenomenon cannot be an origin of spin because
 the one-dimensional Dirac-like Hamiltonian cannot contain spin degree of freedom.
 Further, in contrast to currently unreachable
 length and short time scales  around $10^{-13} \mathrm{m}$
 and  $10^{21} \mathrm{Hz}$, respectively,
 beyond the capability of present technology for
 the Schr\"odinger's Zitterbewegung,
 the quantum simulation of $3+1$ dimensional Dirac theory of
 a trapped ion would reach the experimental feasibility in near future.
 Such a proof-of-principle quantum optical simulation
 of a tunable relativistic quantum mechanical system
 may allow to explore
 the relativistic spin and the proper position operator
 of $3+1$ dimensional spin-$1/2$ Dirac-like particles
 and to provide intriguing and profound understanding of a Dirac particle.

 %
 In a remarkable contrast to previous approaches suggesting relativistic spin operators for spin-$1/2$ massive particles,
 our approach giving the fundamental equation from the relativistic spin operator $\mathcal{S}$ as the generators of the little groups
 in the parity-extended Poincar\'e group
 has enabled to manifest from the Noether's theorem that our relativistic spin $\mathcal{S}$ in equation (\ref{eq:DSP}) is a conserved quantity.
 Also the corresponding spin current $(J^{\mu}_S)^{ij}$ in equation (\ref{eq:ScurrentS}) has been expressed explicitly by using the spin operators
 and has been shown to be conserved.
 The relativistic spin  $\mathcal{S}$
 and spin current $(J^{\mu}_S)^{ij}$ can be naturally extended
 with an electromagnetic interaction and then are applicable
 in exploring future spin-based (relativistic) quantum technologies as well as
 in resolving the inconsistent phenomena, for instance,
 the non-conserving spin current
 due to spin-orbit
 couplings, emerging in spintronics \cite{Rashba,Sun1,Shi,An}

 Our work opens the door for unraveling the puzzling inconsistences in association with
 the fundamentals of spin and its dynamics, and
 for exploring unveiled physics of massive particles for any spin $s$.
 On the more practical level,
 one could use the spin operators
 and the corresponding representation of the Poincar\'e group in exploring
 spin-dependent forces on massive particles.
 On the more fundamental level,
 the fundamental equation, expressed explicitly in terms of the relativistic spin
 as the invariant in the spacetime symmetry, could be considered as a starting point to attempt to set up a quantum gravitational theory
 in the elementary particle domain by exploring how the relativistic spin couples with gravity
 for finding a successful extension of relativistic quantum theory.
%

\section{appendix}


\noindent\textbf{Derivation of spin operator with physical requirements.}
 (i) In classical physics, a spatial angular momentum vector and the total angular momentum
 vector are a spatial three-vector.
 It is then natural that a spin angular momentum vector is also regarded as
 \textit{a spatial three-dimensional vector}.
 Thus, the spin three-vector transforming as a three-dimensional vector under a spatial rotation
 should satisfy
 \begin{equation}
  [ J^j, S^k ] = i\epsilon_{jkl} S^l ,
  \label{eq:rotation}
 \end{equation}
 where $J^j$ is the rotation generator around the axis $\hat x^j$
 and $\epsilon_{jkl}$ is a three-dimensional Levi-Civita with $\epsilon_{123}=1$.
 Let us substitute equation (\ref{eq:linear}) into the commutation relation in equation (\ref{eq:rotation}).
 Equation (\ref{eq:rotation}) becomes
  \begin{equation}
  [ J^j, a_{k,\mu} ] W^\mu + i\epsilon_{jln} a_{k,l}  W^n =  i\epsilon_{jkl} a_{l,\nu} W^\nu
  \label{eq:equal1}
 \end{equation}
 by using
 $J^i = \epsilon_{ijk} J^{jk}/2$ and
 $[ J^{\lambda\mu}, W^\nu ] = i (g^{\mu\nu} W^{\lambda} - g^{\lambda\nu} W^{\mu})$.
 Since all $W^\mu$ terms are linearly independent,
 the coefficients $a_{k,\mu}$ from equation (\ref{eq:equal1}) have the conditions:
\begin{subequations}
 \begin{eqnarray}
 \left[ J^j, a_{k,0} \right] \!\!&=&\!\! i \epsilon_{jkl} a_{l,0}
 \mbox{~\hspace{0.1cm} for~} W^0,
 \label{eq:condition0}
 \\
 \left[ J^j, a_{k,k} \right] \!\! +\!\! i \epsilon_{jlk} a_{k,l} \!\!&=&\!\! i \epsilon_{jkl} a_{l,k}
 \mbox{~\hspace{0.1cm} for~} W^k,
 \label{eq:conditionK}
 \\
 \left[ J^j, a_{k,m \neq k} \right] \!\!+\!\! i \epsilon_{jl(m\neq k)} a_{k,l}
    \!\!&=&\!\! i \epsilon_{jkl} a_{l,m\neq k}
 \mbox{~for~} W^m.
 \label{eq:conditionM}
 \end{eqnarray}
\end{subequations}
 As a function of the momentum operators,
 the coefficient $a_{k,0}$ in equation (\ref{eq:condition0})
 is a function of  $P^k$ and $P^0$ because for $j=k$,
 $[J^k,a_{k,0}]=0$ is guaranteed from
 $[J^k, P^0]=0$ and $[J^k,P^k]=0$
 in the commutation relation between $J^{\mu\nu}$ and $P^\rho$, i.e.,
 $[J^{\mu\nu}, P^\rho] = i(g^{\nu\rho} P^\mu - g^{\mu\rho} P^\nu)$.
 In order to satisfy equation (\ref{eq:condition0}) for $j \neq k$, also,
 $a_{k,0}$ should be a linear function of $P^k$ because if it is a quadratic
 and more higher order function of $P^k$ then
 the left-hand side of equation (\ref{eq:condition0}) becomes zero,
 but the right-handed side of equation (\ref{eq:condition0}) cannot be zero with general momenta.
 Then, the coefficient $a_{k,0}$ of the term $W^0$ can be written as
 \begin{equation}
  a_{k,0} = f_0(P^0) \ P^k,
 \label{eq:ak0}
 \end{equation}
  where $f_0(P^0)$ is a function of $P^0$.


 Equation (\ref{eq:conditionM}) becomes
 $[ J^k, a_{k,m\neq k} ] = i\epsilon_{kml} a_{k,l}$ for $j =k$
 and $[ J^m, a_{k,m\neq k} ] = i\epsilon_{mkl} a_{l,m}$ for $j=m$.
 This implies that
 the non-commuting part of the operator $a_{k,m\neq k}$
 transforms as the $m$- or $k$-component of a three-vector under a rotation.
 In three-dimension, only two types of vectors are possible.
 One is an ordinary vector $\mathbf{P}$,
 the other is a pseudovector $\mathbf{P}\times\mathbf{C}$
 with a constant vector $\mathbf{C}$.
 To satisfy equation (\ref{eq:conditionM}), then, the $a_{k,m\neq k}$ is expressed as
 \begin{equation}
  a_{k,m\neq k} = f_2(P^0) \ P^k P^m + f_3(P^0) \epsilon_{kml} P^l,
 \label{eq:akm}
 \end{equation}
  where $f_2(P^0)$ and $f_3(P^0)$ are functions of $P^0$.


 The coefficient $a_{k,k}$ in equation (\ref{eq:conditionK})
 is a function of $P^k$ and $P^0$ because $a_{k,k}$ commutes with $J^k$ for $j=k$.
 For $j \neq k$, furthermore,
 equation (\ref{eq:conditionK}) becomes $[ J^j, a_{k,k} ] =0$ by using the coefficient
 $a_{k,m\neq k}$ in Eq. (\ref{eq:akm}).
 At this stage, then, the coefficient $a_{k,k}$ is not specified more.
 However, for $j \neq k \neq m$, equation (\ref{eq:conditionM})
 can be $[J^j, a_{k,m\neq k} ] + i\epsilon_{jkm} a_{k,k}  = i\epsilon_{jkm} a_{m,m}$.
 Satisfying this condition,
 $a_{k,k}$ can have a form of $f_1(P^0)$ or $f_2(P^0) P^k P^k$.
 The coefficient $a_{k,k}$ can then be written as
 \begin{equation}
  a_{k,k} = f_1(P^0) + f_2(P^0) P^k P^k,
 \label{eq:akk}
 \end{equation}
  where $f_1(P^0)$ is a function of $P^0$.
 Consequently, as a three-dimensional vector satisfying equation (\ref{eq:rotation}),
 equation (\ref{eq:linear})
 can be rewritten in terms of a more specific form of the coefficients $a_{k,\mu}$:
 \begin{eqnarray}
 S^k = f_0(P^0) P^k W^0 + f_1(P^0) W^k + f_2(P^0) P^k P^n W^n
        + \ f_3(P^0)\ \epsilon_{kml}\ P^l W^m.
 \label{sol1}
 \end{eqnarray}


 (ii) The spin three-vector operators are
 \textit{generators of $\mathit{SU}(2)$ group}
  such that
  they should satisfy the $\mathfrak{su}(2)$ algebra, i.e., the commutation relations,
  \begin{equation}
  [ S^i, S^j ] = i\epsilon_{ijk} S^k.
  \label{eq:su2}
  \end{equation}
  Let us put equation (\ref{sol1}) into the commutation relation in equation (\ref{eq:su2}).
  By using the commutation relations $[W^0, W^k] = i\epsilon_{klm} W^l P^m $
  and $[W^i,W^m] = i\epsilon_{iml} (W^l P^0  - W^0 P^l)$,
  three equations are obtained as
\begin{subequations}
 \begin{eqnarray}
 f_0 + f_2 P^0 \!\!&=&\!\! - f_0 f_1 P^0 - f_1^2 + m^2 f_3^2 - f_1 f_2 P_0^2,
 \label{eq:f1}
 \\
 f_1 \!\!&=&\!\! (f_0 + f_2)\ f_1\ (P_0^2 - m^2) +\ f_1^2\ P^0,
 \label{eq:f2}
 \\
 f_3 \!\!&=&\!\! (f_0 + f_2)\ f_3\ (P_0^2 - m^2) + f_1 f_3 P^0.
 \label{eq:f3}
 \end{eqnarray}
\end{subequations}
 From equations (\ref{eq:f1}), (\ref{eq:f2}), and (\ref{eq:f3}), however,
 $f$'s cannot be determined because the three equations have the four variables,
 which means that infinitely many solutions are possible with respect to $f$'s.
 Moreover, equations (\ref{eq:f2}) and (\ref{eq:f3}) are not independent each other.


 (iii) To specify $f$'s more, we consider the fact that the angular momentum three-vectors
 are obtained from the second-rank tensors.
 In the same manner of
 the relation between the angular momentum tensor and the angular momentum three-vector,
 the spin three-vector is denoted by using the spatial components of a spin tensor $S^{\mu\nu}$,
  i.e.,
  \begin{equation}
  S^k = \frac{1}{2} \epsilon_{klm} S_{lm}.
  \label{eq:tensor}
  \end{equation}
 Crucially, $S^k$ is the $k0$-component of the dual spin tensor
 $ {}^*S^{\mu\nu}=\frac{1}{2}\epsilon^{\mu\nu\rho\sigma} S_{\rho\sigma}$, i.e.,
 \begin{equation}
  {}^*S^{k0} = S^k
  \label{eq:dualtensor}
 \end{equation}
 because ${}^*S^{k0}=\frac{1}{2}\epsilon^{k0lm}S_{lm}
 =\frac{1}{2}\epsilon_{klm}S_{lm}$.
 Hence, equations (\ref{eq:tensor}) and (\ref{eq:dualtensor})
 imply that
 $S^k$ should be transformed as
 a $k0$-component of a second-rank tensor for a LT.
 In fact, this tensorial requirement
 is a generalization of the spatial three-vector condition in equation (\ref{eq:rotation})
 because if a LT becomes a spatial rotation
 then the tensorial requirement reduces to the spatial three-vector condition in (i).
 As well as the spatial three-vector condition in equation (\ref{eq:rotation}),
 thus, the tensorial requirement gives the additional condition.

 Then, $f$'s given in equations (\ref{eq:ak0}), (\ref{eq:akm}), and (\ref{eq:akk})
 from the spatial three-vector condition can be specified more by the additional condition
 as follows.
 Under a LT,
 $f_1$ should be linearly proportional to $P^0$, i.e., $f_1(P^0) = b\, P^0$,
 to make the term of $f_1(P^0)W^k$ transforming like a $k0$-component of the tensor,
 while $f_0$ and $f_3$ should be constant (scalar), i.e., $f_0(P^0) = a$
 and $f_3(P^0) = c$, because, for instance, the terms of $P^k W^0$ and
 $\epsilon_{kml}P^l W^m = \epsilon_{0kml} P^l W^m$ already transform like
 a $k0$-component, where $b$ and $c$ are constant (scalar under the LT).
 However, the term of $f_2(P^0) P^k P^n W^n$ is converted to a form of
 $f_2(P^0) P^k W^0 P^0$ by using $W^\mu P_\mu=0$.
 This implies that actually this term
 is a $k00$-component of a third-rank tensor.
 Thus, to satisfy the tensorial property of the second-rank spin tensor,
 one has to set $f_2(P^0)=0$.
 Consequently,
 equation (\ref{sol1}) can be rewritten as a more specific form:
 \begin{equation}
 S^k = a\ P^k W^0 + b\ P^0 W^k + c\ \epsilon_{kml} P^l W^m.
 \label{newS}
 \end{equation}

 On substituting equation (\ref{newS}) into equation (\ref{eq:su2}),
 equations (\ref{eq:f1}), (\ref{eq:f2}), and (\ref{eq:f3}) become,
 respectively,
\begin{subequations}
 \begin{eqnarray}
 a  &=& - a\ b\ P_0^2 - b^2 P_0^2 + m^2\ c^2,
 \label{eq:f1a}
 \\
 b &=& a\  b\ (P_0^2 - m^2) + b^2\ P_0^2,
 \label{eq:f2b}
 \\
 c &=& a\  c\ (P_0^2 - m^2) + b\ c\ P_0^2.
 \label{eq:f3c}
 \end{eqnarray}
\end{subequations}
 For an arbitrary $P^0$,
 the three equalities in equations (\ref{eq:f1a}), (\ref{eq:f2b}) and (\ref{eq:f3c}) should hold,
 which means that both the coefficients of $P^0$
 and the constant terms in the equalities should be zero.
 To determine the three constants $a$, $b$, and $c$, then,
 we obtain the six conditions:
\begin{subequations}
 \begin{eqnarray}
 && a\ (a + b) = 0 \mbox{~~~and~~~~~} a\ - m^2 \ c^2 =0,
 \\
 && b\ (a + b) = 0 \mbox{~~~and~~} b\ ( 1 + m^2\ a) = 0,
 \\
 && c\ (a + b) = 0 \mbox{~~~and~~~}  c\ ( 1 + m^2\ a) = 0.
 \end{eqnarray}
\end{subequations}
 These six conditions clearly show that if one of the three constants is zero then
 all of the three constants become zero.
 Hence, all of them should be nonzero and then
 the six conditions reduce to
 the three conditions:
\begin{subequations}
 \begin{eqnarray}
  a + b &=& 0,
  \label{eq:three1}
 \\
  1 + m^2\ a  &=& 0,
 \label{eq:three2}
 \\
  a - m^2\ c^2 &=& 0.
  \label{eq:three3}
 \end{eqnarray}
\end{subequations}
 One can obtain the two sets of the three constants as
 \begin{equation}
 a =-\frac{1}{m^2}, \mbox{~~~} b = \frac{1}{m^2}, \mbox{~~and~~} c = \pm \frac{i}{m^2}.
 \end{equation}
 Resultantly, we obtain the two spin three-vectors as
 \begin{equation}
 S^k_{\pm} = \frac{1}{m^2} \left( P^0 W^k - P^k W^0\right) \pm \frac{i}{m^2} \epsilon_{kml} P^l W^m .
 \nonumber
 \end{equation}
 Note that in deriving the two spin operators in equation (\ref{eq:pm}),
 we used the minimal conditions, i.e.,
 the $\mathfrak{su}(2)$ algebraic requirement in (ii)
 and the tensorial requirement in (iii),
 because the tensorial requirement includes
 the spatial three-vector condition in (i).
%

%
 \textit{Transformation operator.}
 The spin operator $S^k_\pm (\mathbf{p})$ in equation (\ref{eq:spin}) have been obtained
 from the $S^k_\pm$ in Eq. (\ref{eq:pm}) by using the standard boost LT $L(\mathbf{p})$
 from the rest frame to the moving frame with the momentum $\mathbf{p}$.
 This implies that
 the spin operator $S^k_\pm (\mathbf{p})$ in equation (\ref{eq:spin})
 can be reexpressed as a usual transformation form, i.e.,
 $ U_\pm S^k_\pm (\mathbf{0}) U^{-1}_\pm$ in terms of
 the rest spin operator $S^k_\pm (\mathbf{0})$
 with a transformation operator $U$.
 Prior to manipulate the right-handed side of equation (\ref{eq:spin}),
 let us define
 $\cosh \frac{\xi}{2} = \sqrt{\frac{p^0+m}{2m}}$ and $\sinh \frac{\xi}{2} = \sqrt{\frac{p^0-m}{2m}}$
 with
 $(p^0)^2 = |\mathbf{p}|^2 + m^2$.
 One can then manipulate the right-handed side of  $S^k_+(\mathbf{p})$ in Eq. (\ref{eq:spin}) such as
\begin{subequations}
 \begin{eqnarray}
  S^k_+ (\mathbf{p})
  \!\!&=&\!\! \frac{\sigma^k}{2} + \sinh \xi  \ A^k
        + (\cosh \xi -1) \ B^k
 \\
 \!\!&=&\!\! \frac{\sigma^k}{2} + \sum_{n=1}
    \left[ \frac{\xi^{2n-1}}{(2n-1)!}\    A^k
        + \frac{\xi^{2n}}{2n!}  \ B^k
         \right],
  \label{eq:series}
 \end{eqnarray}
\end{subequations}
 where $A^k = i\left(\mbox{\boldmath $\sigma$} \times \hat{\mathbf{p}}\right)^k/2$
 and $B^k = \sigma^k/2- \hat{p}^k \left(\mbox{\boldmath $\sigma$} \cdot \hat{\mathbf{p}} \right)/2$.
 One can notice
 that equation (\ref{eq:series}) can be expressed as a form of
 $e^X Y e^{-X} = Y + \sum_{n=1} X_n/n!$ with $X_{n+1}=\frac{1}{n+1}[X, X_n]$ and $X_1=[X, Y]$
 in the Baker-Hausdorff formula because the first term can be $Y=\sigma^k/2$.
 Then, let us work out an explicit form of the operator $X$ by assuming
 the transformation operator as $U_+ = \exp[ X ]$,
 where $X=f(\mbox{\boldmath $\sigma$})$ is a function of the rest spin operator
 $\mbox{\boldmath $\sigma$}$ with the given momentum $p^\mu$ in a moving frame.
 In terms of the function $f(\mbox{\boldmath $\sigma$})$, the recursive relation is given as
 $X_{n+1} = \frac{1}{n+1} [ f(\mbox{\boldmath $\sigma$}), X_{n}]$ with
 $X_1 = [ f(\mbox{\boldmath $\sigma$}), \sigma^k/2]$.
 Comparing with equation (\ref{eq:series}),
 we have the two relations
 $X_{2n-1} = \frac{\xi^{2n-1}}{(2n-1)!}
                  \left(\mbox{\boldmath $\sigma$}/2 \times \hat{\mathbf{p}}\right)^k $
 and
 $
 X_{2n} = \frac{\xi^{2n}}{2n!}  \left( \sigma^k/2
        - \hat{p}^k \left(\mbox{\boldmath $\sigma$}/2 \cdot \hat{\mathbf{p}}\right)\right).
 $
 In determining the function $f(\mbox{\boldmath $\sigma$})$, thus, we have the two conditions
 $X_1 = [ f(\mbox{\boldmath $\sigma$}), \sigma^k/2]
 = \xi \left(\mbox{\boldmath $\sigma$}/2 \times \hat{\mathbf{p}}\right)^k $
 and
 $   \frac{1}{2n+1} [ f(\mbox{\boldmath $\sigma$}), \left(\mbox{\boldmath $\sigma$}/2 \times \hat{\mathbf{p}}\right)^k ]
  = \frac{\xi}{2n}  \left( \sigma^k/2
        - \hat{p}^k \left(\mbox{\boldmath $\sigma$}/2 \cdot \hat{\mathbf{p}}\right)\right).$
 By using the $\mathfrak{su}$(2) algebra $[\sigma^i, \sigma^j] = 2 i \epsilon_{ijk}\sigma^k$,
 we see
 $\left(\mbox{\boldmath $\sigma$} \times \hat{\mathbf{p}}\right)^k
 = [ \sigma^j, \sigma^k] \hat{\mathbf{p}}^j/2 $
 and then find $f(\mbox{\boldmath $\sigma$}) =\xi\, \sigma^j \hat{\mathbf{p}}^j/2$.
 By putting the function $f(\mbox{\boldmath $\sigma$}) =\xi\, \sigma^j \hat{\mathbf{p}}^j/2$ into
 the second condition, one can find that the equality of the second condition holds.
 The $S^k_+(\mathbf{p})$ in equation (\ref{eq:spin}) is reexpressed as
\begin{equation}
  S_+(\mathbf{p})
  =
  \exp\!\Big[\frac{1}{2} \mbox{\boldmath $\sigma$}\cdot \mbox{\boldmath $\xi$} \Big]
  \left( \frac{\sigma^k}{2}  \right)
  \exp\!\Big[-\frac{1}{2} \mbox{\boldmath $\sigma$}\cdot \mbox{\boldmath $\xi$} \Big].
 \nonumber
\end{equation}
 Consequently,
 the spin oprator $S^k_+(\mathbf{p})$
 is the standard boost LT of the rest spin operator $S^k_+(\mathbf{0})$
 and the LT operator for the spin operators is defined as
\begin{equation}
  U_+= \exp\left[ \frac{\mbox{\boldmath $\sigma$}}{2}\cdot \mbox{\boldmath $\xi$} \right].
  \nonumber
\end{equation}
 Similarly, we also obtain
 $U_-= \exp\left[- \mbox{\boldmath $\sigma$}/2\cdot \mbox{\boldmath $\xi$} \right] $
 from the $S^k_-(\mathbf{p})$ in equation (\ref{eq:spin}).
%

%

\noindent\textbf{Little groups.}
 Since the group elements $\mathcal{D}_\pm(\theta^k_\pm)$ are
 respectively generated by the spin operators $S^k_\pm({\bf p})$,
 the representation spaces of these groups are
 composed of the eigenstates $\Psi_\pm(p^\mu,\lambda)$.
 To study the case that gives the little group in these representation spaces,
 let us first consider the non-collinear successive Lorentz transformations,
 transforming the particle back to its rest frame, from the PRF.
 All non-collinear successive Lorentz transformations $\Lambda$ and $L(\mathbf{p})$
 with $\Lambda L(\mathbf{p}) \neq L(\Lambda  \mathbf{p})$ can give the effective rotation in the PRF:
 \begin{equation}
   R(\Lambda,  \mathbf{p})= L^{-1}(\Lambda  \mathbf{p}) \Lambda L(\mathbf{p}),
 \label{eq:STWR}
 \end{equation}
 where $L^{-1}(\Lambda \mathbf{p})$ is the inverse of $L(\Lambda \mathbf{p})$,
 $\Lambda L(\mathbf{p})$ transforms the rest momentum $k^\mu$ to
 $q^\mu=\Lambda^\mu_{\phantom{\mu}\nu}L(\mathbf{p})^\nu_{\phantom{\rho}\rho}k^\rho $,
 and
 $L(\mathbf{q})$ is a standard Lorentz transformation giving
 $q^\mu=L(\Lambda  \mathbf{p})^\mu_{\phantom{\rho}\rho}k^\rho$,
 because
 $\Lambda L(\mathbf{p})$ is equivalent
 to the rotation followed by the standard LT $L(\Lambda  \mathbf{p})$, i.e.,
 $\Lambda L(\mathbf{p}) = L(\Lambda  \mathbf{p}) R(\Lambda,  \mathbf{p})$.
 In the representation space composed of the eigenstates $\Psi(k^\mu,\lambda)$
 at the PRF, the rotation $R(\Lambda,  \mathbf{p})$ is represented by
$\tilde{\mathcal{D}} (R(\Lambda,\mathbf{p}))=\exp[{\frac{i}{2} \phi^k(\Lambda,\mathbf{p})\, \sigma^k}]$,
where the rotation angle $ \phi^k$
 are determined by $\Lambda$
 and $L(\mathbf{p})$ through the rotation $R(\Lambda, \mathbf{p})$.
One can see that the representation of $\tilde{\mathcal{D}}_\pm (R(\Lambda,\mathbf{p}))$ is
 nothing but the element $\mathcal{D}_\pm(\theta^k_\pm) $ generated by the
 $S^k_\pm (\mathbf{0}) = \sigma^k/2$
 with the angle $\theta^k_\pm = \phi^k (\equiv \theta^k)$ because
$\mathcal{D}_\pm(\theta^k_\pm)$ in the PRF has nothing to do with the handedness.
 The spin state representation of the rotation $R(\Lambda, {\bf p})$, i.e.,
 $ U^{-1}_\pm [L({\Lambda \mathbf{p}})] U_\pm [\Lambda] U_\pm [L(\mathbf{p})]$,
 acting on the spin state space in the PRF
 can be written as
\begin{eqnarray}
 \tilde{\mathcal{D}}_\pm (R(\Lambda,\mathbf{p}))
 =U_\pm [L^{-1}(\Lambda  \mathbf{p}) \Lambda L(\mathbf{p})]
 \label{littleL}
\end{eqnarray}
 with using the group laws $U[A]U[B]U[C]=U[ABC]$ and $U^{-1}[A]=U[A^{-1}]$.
 In the moving reference frame with the particle momentum
 $q^\mu = L( \mathbf{q})^\mu_{\phantom{\rho}\rho}k^\rho$,
 the little group representation can then be obtained such as
\begin{equation}
  U_\pm[L(\mathbf{q})]
  \tilde{\mathcal{D}}_\pm (R(\Lambda,\mathbf{p})) U^{-1}_\pm[L(\mathbf{q})]
  \Psi_\pm (q^\mu,\lambda)
  =   \exp\!\Big[{\frac{i}{2} \theta^k S^k_\pm}(\mathbf{q})\Big]
  \Psi_\pm(q^\mu,\lambda),
 \label{eq:little}
\end{equation}
 where
 the spin state are represented
 by $\Psi_\pm(q^\mu,\lambda)= U_\pm[L(\mathbf{q})] \Psi(k^\mu,\lambda)$
 with the standard LT $U_\pm[L(\mathbf{q})]$.
 In equation (\ref{eq:little}),
 the $\exp[ \frac{i}{2} \theta^k S^k_\pm (\mathbf{q})]$ is
 the elements $\mathcal{D}_\pm(\theta^k) $ of the $\mathrm{SU}(2)$ groups
 generated by $S^k_\pm (\mathbf{q})$ in the reference frame.
  Hence
  the group element $\mathcal{D}_\pm(\theta^k)$ is determined solely by the representations of
  the LTs, $U_\pm[\Lambda]$, $U_\pm[L(\mathbf{p})]$,
 and
 $U_\pm[L(\mathbf{q})]$.
 Therefore, the two $\mathrm{SU}(2)$ groups with the elements $\mathcal{D}_\pm(\theta^k)$
 generated by the $S^k_\pm(\mathbf{q})$ in the reference frame
 are the little groups.
 Their elements rotate the spin states without change of
 the momentum eigenvalue of the spin state.
 The rotation angle $\theta^k$ are determined from
 the detailed information of successive Lorentz transformations that give
 the particle's momentum $\mathbf{q}$.
 One of the two little groups can be chosen to represent the Poincar\'e group.

 In addition, equation (\ref{eq:little}) can be written as
\begin{eqnarray}
 \exp\!\!\Big[\frac{i}{2} \theta^k S^k_\pm(\mathbf{q})\Big]\! \Psi_\pm(q^\mu,\lambda)
 \!=\!
 U_\pm [L(\mathbf{q})] \exp\!\!\Big[{\frac{i}{2} \theta^k \sigma^k}\Big]\!
 \Psi(k^\mu,\lambda).
 \label{eq:13}
\end{eqnarray}
 Note that for both rotations on the spin states in the PRF (the right-handed side of equation (\ref{eq:13}))
 and in the moving frame (the left-handed side of equation (\ref{eq:13})),
 the angle parameter $\theta^k$ does not change.
 Also,
 the spin eigenvalues of $S^k_\pm(\mathbf{p})$
 and $\sigma^k$ are the same from equations (\ref{eq:covariance}) and (\ref{eq:TSPIN}).
 These imply that all three rotations, respectively generated
 by the spin operators $S^k_+({\bf p})$, $S^k_-({\bf p})$, and $\sigma^k$, are equivalent.
 Consequently,
 for the spin state,
 the two spin operators $S^k_\pm({\bf p})$ in the arbitrary frame in equation (\ref{eq:13})
 provide the same little group rotation.
%

%
\noindent\textbf{Covariant parity operator.}
 The covariant parity operator in the spin space
 should be represented by using the spin operators
 $S^k_\pm(\mathbf{p})$ and momentum $p^\mu$.
 It is expected that the covariant parity operator
 would be a $0$-th component of four-vector because
 the usual representation of the parity operator
 is $\gamma^0$ and
 $\bar \psi_D(p^\mu,\lambda)
 \gamma^0 \psi_D(p^\mu,\lambda)$
 in the four-dimensional representation is
 the $0$-th component of four-vector \cite{Schwartz}.
 Hence the most desirable candidate has a form of
 ${\cal P} =a p^0 + b\, {}^*S_{0\mu;\pm}(\mathbf{p})\, p^\mu$,
 where $a$ and $b$ are Lorentz-invariant coefficients,
 because
 the operators $^*S_{0\mu;\pm}(\mathbf{p})\, p^\mu = S^k_\pm (\mathbf{p})\,
 p^k$
 become $w^0$ from equation (\ref{eq:pm})
 and
 then $S^k_\pm\, p^k =\sigma^k\, p^k/2$ due to
 $w^0 = \boldsymbol{\sigma}/2\cdot \mathbf{p}$ in an arbitrary frame.
 To find such a desirable form of covariant parity operator,
 let us consider the two consecutive LT operations  $U^{\, 2}_\mp [ L(\mathbf{p})]$
 on the states $\Psi_\pm (p^\mu,\lambda)$,
 i.e., $U^{\, 2}_\mp [ L(\mathbf{p})] \Psi_\pm (p^\mu,\lambda)$.
 Since the relations
 $U^{-1}_\pm[ L(\mathbf{p})] = U_\mp[ L(\mathbf{p})] $ in equation
 (\ref{eq:LT}) and
 $\Psi_\pm (p^\mu,\lambda) = U_\pm [ L(\mathbf{p})]
 \Psi (k^\mu,\lambda)$ in equation (\ref{eq:TSPIN}),
 the two consecutive LT operations
 convert the states $\Psi_\pm (p^\mu,\lambda)$
 into the states $\Psi_\mp (p^\mu,\lambda)$, i.e.,
\begin{equation}
  U^{\, 2}_\mp [ L(\mathbf{p})]\, \Psi_\pm (p^\mu,\lambda) = \Psi_\mp (p^\mu,\lambda).
\end{equation}
 This means that for any spin, the two consecutive LT
 operators $U^{\, 2}_\pm [ L(\mathbf{p})]
 = \exp\left[\pm \mbox{\boldmath $\sigma$} \cdot \mbox{\boldmath $\xi$} \right]$
 can be parity operators for the state $\Psi_\mp (p^\mu,\lambda)$, respectively.
 If one considers the Taylor series expansion of the two consecutive LT operators,
 with the condition $\{\sigma^i,\sigma^j\}= \sigma^i\sigma^j + \sigma^j
 \sigma^i = 2\delta^{ij}$
 in the terms including
 $\left(\boldsymbol{\sigma} \cdot \boldsymbol{\xi} \right)
 \left(\boldsymbol{\sigma} \cdot \boldsymbol{\xi} \right)
 = \{\sigma^i,\sigma^j\} \xi^i \xi^j/2$,
 one can have
\begin{equation}
 U^{\, 2}_\pm [L(\mathbf{p})] = \cosh |\boldsymbol{\xi}| \pm
 \frac{\boldsymbol{\sigma}\cdot \mathbf{p}}{|\mathbf{p}|} \sinh
 |\boldsymbol{\xi}|,
 \label{eq:LT2}
\end{equation}
 where $\mathbf{p} \parallel \boldsymbol{\xi}$ has been used.
 Since $\boldsymbol{\sigma}\cdot \mathbf{p} = \mathbf{S}_\pm \cdot
 \mathbf{p}$,
 equation (\ref{eq:LT2}) can be covariant and
 has the desirable form for covariant parity operators.
 In consequence, the covariant parity operator
 is represented as
\begin{equation}
 \frac{1}{m}\left( p^0 \pm {}^* S_{0\mu}\, p^\mu \right)
  \Psi_\pm (p^\mu,\lambda) = \Psi_\mp (p^\mu,\lambda).
\label{covariantP}
\end{equation}
 However, the condition $\{\sigma^i,\sigma^j\} = 2 \delta^{ij}$
 is satisfied only for spin-$1/2$ representation.
 Then, only for spin-$1/2$ case, the parity operator can be covariantly
 represented in equation (\ref{covariantP}).
 In the direct sum $(1/2,0)\oplus (0,1/2)$ representation,
 the covariant parity operator can be rewritten as
 $\mathcal{P} = ({P^0 + \Sigma^k P^k \gamma^5})/{m}$.

 \underline{~\hspace*{4cm}}

\section*{Acknowledgements} 
We acknowledge support from the National Research Foundation of Korea Grant funded by the Korea Government (2015-0226, T.C.)
 and the National Natural Science Foundation of China under the Grant
 No. 11374379 (S. Y. C.).


\end{document}